\documentstyle[aps,prl,preprint,epsf]{revtex}
\begin{document}

\draft
\title{Femtosecond Coherent Dynamics of the Fermi Edge 
Singularity and Exciton  Hybrid}
\author{T. V. Shahbazyan, N. Primozich,
and I. E. Perakis}
\address{Department of Physics and Astronomy,  
Vanderbilt University, Nashville, TN 37235}
\author{D. S. Chemla}
\address{Department of Physics,  
University of California, Berkeley, CA 94720
and Materials Science Division, Lawrence Berkeley National
Laboratory,  Berkeley, CA 94720
}
\maketitle

\begin{abstract}
We study theoretically the coherent nonlinear optical
response of doped semiconductor quantum wells with several
subbands. When the Fermi energy 
approaches the exciton level of an upper subband, the
absorption spectrum acquires a characteristic double--peak shape
originating from the interference between the Fermi--edge
singularity and the exciton resonance. We demonstrate that, for 
off--resonant pump excitation, the pump/probe spectrum
undergoes a striking transformation in the coherent regime,
with a time--dependent exchange of oscillator strength
between the Fermi edge singularity and exciton peaks. We show
that this effect originates from the many--body electron--hole
correlations which determine the dynamical response of
the Fermi sea. Possible
experimental applications are discussed. 
\end{abstract}

\maketitle
\narrowtext

Ultrafast nonlinear spectroscopy offers a
unique perspective into the role of many--body effects in
semiconductors\cite{shah}. While the linear absorption
spectrum provides information about static properties, 
ultrafast time--resolved 
spectroscopy allows one to probe the system on time scales
shorter than those governing the interactions between
the elementary excitations. 
In the coherent regime, the dynamics
of many--body correlations plays an important role in
the transient changes of the absorption spectrum\cite{axt98}. 
For example, in undoped semiconductors, exciton--exciton interactions
were shown to play a dominant role in the optical response for
specific sequences of the optical pulses\cite{kne98,ost95,che98}. 

In modulation--doped quantum wells (QW), the optical properties
are dominated by the Fermi edge singularity (FES)\cite{mahan}. 
Unlike in the undoped case, where the linear absorption exhibits 
discrete bound state peaks whose width is ultimately determined by the
homogeneous broadening, the FES is a continuum resonance whose
lineshape is governed by the Coulomb interactions of the 
photoexcited carriers with the low--lying Fermi sea (FS) excitations. 
In this letter we study the
role of such many--body correlations in pump/probe
measurements, where the  strong ``pump'' pulse excites the system
at time $t=0$,  while the weak ``probe''
pulse measures the optical response at time $t=\tau$. 
In the  doped systems, the
interactions are screened and there are no discrete bound states. 
Therefore, the many--body correlations enter into the nonlinear
response not via exciton--exciton interactions, but
mainly through the dynamical response of the FS 
during the course of the optical excitation. Since such electron--hole
({\em e--h}) correlations come from the ``dressing'' of the 
photoexcited {\em e--h} pair by {\em gapless} FS excitations, the
response of the FS to ultrashort optical pulses is intrinsically
{\it unadiabatic} \cite{per94,pri99}.
For resonant pump excitation, the electron--electron 
({\em e--e}) scattering also leads to  strong variations of the
dephasing and relaxation times, from  a few ps 
close to the FES to $\sim 10$ fs away from the Fermi 
level\cite{knox88,kim92,wang95}. However, such incoherent 
relaxation effects are  suppressed for {\em off}--resonant
excitation, when the pump is tuned below the FES resonance,
in which case coherent effects dominate\cite{bre95,pri99}.
Furthermore, for {\em negative} ($\tau <0$) time delays, the
pump/probe signal is due to the coherent interaction of the
pump pulse with the polarization induced in the sample
by the probe pulse and, again, the effects of incoherent
processes are strongly attenuated \cite{fluegel87}.
Therefore, the {\em coherent} dynamics studied below
can be best observed under off--resonant conditions or
when the probe precedes the pump. 

Here we investigate the ultrafast pump/probe dynamics of the
FES--exciton hybrid, which is formed in asymmetric QW's with
partially occupied subbands\cite{chen90,mue90,sko91}. 
In such structures, interband optical transitions 
from the  valence band to several conduction subbands are allowed
due to the finite overlap between the hole and electron envelope
wave--functions. The many--body effects on the linear absorption
spectrum have been described by using the simple two--subband
Hamiltonian\cite{mue90} 
\begin{equation}
\label{H0}
H=H_0
-\sum_{ij}\sum_{\bf pkq}v_{ij}({\bf q}) a_{i\bf p+q}^{\dagger}a_{j\bf p}
b_{\bf k-q}^{\dagger}b_{\bf k},
\end{equation}
with 
$H_0=\sum_{i\bf k}\epsilon_{i\bf k}^ca_{i\bf k}^{\dagger}a_{i\bf k}+
\sum_{\bf k} (\epsilon_{\bf k}^v+E_g)b_{\bf -k}^{\dagger}b_{\bf -k}$.
Here $a_{i\bf k}^{\dagger}$ and $\epsilon_{i\bf k}^c$
are the creation operator and the energy of a conduction
electron in the $i$th subband, $b_{\bf -k}^{\dagger}$ and 
$\epsilon_{\bf k}^v$ are those of a valence hole
($E_g$ is the bandgap), and $v_{ij}({\bf q})$ is the {\em screened}
{\em e--h} interaction matrix with diagonal (off--diagonal)
elements describing the intrasubband (intersubband)  scattering. Due
to the screening, the interaction potential is short--ranged
and can be replaced by its s--wave 
component \cite{s-r}; close to the Fermi surface, 
$v_{ij}({\bf q})\simeq v_{ij}$ \cite{mue90,mahan}.
Here we consider the case where only the first subband is occupied, but
the Fermi level is close to the exciton level (with binding energy $E_B$) 
below the bottom of the second subband [see inset in Fig.\ 1(a)].
For large values of the FES--exciton splitting
$\Delta-E_F-E_B$, where $\Delta$ is the subband separation, 
the linear absorption spectrum consists of two well separated peaks,
the lower corresponding to the FES from subband 1, and the higher
corresponding to the Fano resonance from the exciton of subband 2
broadened by its coupling to the continuum of states in subband 1. With
decreasing  $\Delta-E_F-E_B$, the FES and the exciton become hybridized 
due to the intersubband scattering arising from the Coulomb interaction.
This results in the transfer of oscillator strength from the exciton
to the FES and a strong enhancement of the absorption peak near the
Fermi level due to the resonant scattering of the photoexcited
electron by the exciton level\cite{chen90,mue90}.

In typical QW's, $v_{12}$ is much smaller than $v_{ii}$  
(a value $v_{12}/ v_{11}\simeq 0.2$ was deduced from the 
fit to the linear absorption spectrum in \cite{mue90}).
In the absence of coupling ($v_{12}=0$),
the different nature of the exciton  and FES
leads to distinct dynamics under  ultrafast excitation\cite{pri99}.
In the presence of coupling, one should expect new
effects coming from the interplay of this difference 
and the intersubband
scattering that hybridizes the two resonances. Indeed, 
we demonstrate that, at negative time
delays, the pump/probe spectrum undergoes a drastic transformation
due to a transient light--induced redistribution of the oscillator
strength between the FES and the exciton. We show that
such a redistribution is a result of the dynamical FS response to
the pump pulse. In fact, the ultrafast pump/probe spectra of
the FES--exciton hybrid can serve as an experimental test of the
difference between the FES and exciton dynamics. 

{\em Theory}.--- The total Hamiltonian of the system is 
$H+H_{p}(t)+H_{s}(t)$, with $H_{\alpha}(t)$ 
($\alpha=p,s$) describing the optical excitations, 
\begin{equation}
\label{opt}
H_{\alpha}(t)=-{\cal E}_{\alpha}(t)
\sum_{i}\left[\mu_iU_i^{\dag}e^{-i\omega_p t+i{\bf k}_{\alpha}{\bf r}}
+ \mbox{h.c.}\right],
\end{equation}
where 
$U_i^{\dag}=\sum_{\bf k}a_{i\bf k}^{\dagger}b_{i \bf -k}^{\dagger}$
is the transition operator to the $i$th subband, $\mu_i$ is the
dipole matrix element, and ${\cal E}_{\alpha}(t)$ are the amplitudes
of the probe ($\alpha=s$) and  pump ($\alpha=p$) electric fields,
propagating in the directions ${\bf k}_{\alpha}$. 
In order to account for the {\em e--h} correlations that
govern the dynamics of the hybrid, we use the multi--subband
generalization of the method developed previously for the
FES\cite{pri99,per94}. The non--linear pump/probe spectrum of the 
system described by the ``bare'' Hamiltonian (\ref{H0}) represents the
linear response to the probe alone of the system described by
the Hamiltonian $H+H_{p}(t)$. In order to take advantage of the
linear--response formalism, we adopt the ``pump--dressed''
effective Hamiltonian $\tilde{H}(t)$, which we derived from
$H+H_{p}(t)$ using a time--dependent Schrieffer--Wolff
transformation\cite{s-w,pri99}. In fact, such a treatment
mimics nicely the spirit of the pump/probe experiments. 
The details will be published elsewhere, and here we present only
the final expressions. 
Since the  pump/probe signal is linear in the probe field, 
the essential physics can be captured by assuming
a $\delta$--function probe pulse,  
${\cal E}_{\tau}(t)={\cal E}_{\tau}e^{i\omega_{p}\tau}\delta(t-\tau)$,
and a Gaussian pump pulse. 
The pump/probe polarization has the form ($t>\tau$)
\begin{equation}
\label{pol}
P(t)=-i{\cal E}_{s}e^{-i\omega_pt}
\sum_{ij}\mu_i\mu_j
\langle 0|\tilde{U}_{i}(t){\cal K}(t,\tau)
\tilde{U}_{j}^{\dag}(\tau)|0 \rangle,
\end{equation}
where ${\cal K}(t,\tau)$ is the time--evolution operator for the
effective Hamiltonian,
\begin{equation}
\label{eff1}
\tilde{H}(t)=\sum_{ij\bf k} \epsilon^c_{ij\bf k}(t)
a_{i\bf k}^{\dagger}a_{j\bf k}+
\sum_{\bf k} \epsilon_{\bf k}^v(t)b_{\bf -k}^{\dagger}b_{\bf -k}
+V_{eh}(t)+
V_{ee}(t),
\end{equation}
where $V_{eh}$ and  $V_{ee}$ are the effective 
{\em e-h} and {\em e-e} interactions and $\tilde{U}_{i}^{\dag}(t)$
is the effective transition operator given below. Here  
$\epsilon^c_{ij\bf k}(t)=\delta_{ij}\epsilon_{i\bf k}^c
+\Delta\epsilon_{ij\bf k}^c(t)$, and
$\epsilon_{\bf k}^v(t)=\epsilon_{\bf k}^v+\Omega+\Delta\epsilon_{\bf k}^v(t)$
are the band dispersions with pump--induced self--energies:
$\Delta\epsilon_{ij\bf k}^c(t)=
-{\cal E}_p(t) [\mu_i p_{j{\bf k}}^{\ast}(t)+
\mu_j p_{i{\bf k}}(t)]/2$,
and
$\Delta\epsilon_{\bf k}^v(t)=
-{\cal E}_p(t){\rm Re}\sum_i\mu_ip_{i{\bf k}}(t)$, 
with $p_{i{\bf k}}(t)$ satisfying
\begin{equation}
\label{sigmai}
i\frac{\partial p_{i{\bf k}}(t)}{\partial t}=
(\epsilon_{i\bf k}^c+\epsilon_{\bf k}^v+\Omega)
p_{i{\bf k}}(t)
-\sum_{j{\bf q}}v_{ij}p_{j{\bf q}}(t)-\mu_i{\cal E}(t),
\end{equation}
where $\Omega$ is the detuning 
of $\omega_{p}$ measured from the Fermi level \cite{pri99}.
Since $p_{i{\bf k}}(t)$ are linear in ${\cal E}_p(t)$, the
self--energies are quadratic in the pump field. Note that
(i) the time--dependence of the self--energies lasts for the
duration of the pump, and (ii) that the pump induces additional
intersubband scattering, described by $\Delta\epsilon_{12\bf k}^c(t)$. 
The effective transition operator appearing in Eq.\ (\ref{pol}) is
$\tilde{U}_i^{\dag}(t)=\sum_{j{\bf k}}\phi_{ij{\bf k}}(t)
a_{j\bf k}^{\dagger}b_{\bf -k}^{\dagger}$,
with 
\begin{equation}
\label{phiij}
\phi_{ij{\bf k}}(t)
=\delta_{ij}\left[1-\frac{1}{2}\sum_l|p_{l{\bf k}}(t)|^2\right]
-\frac{1}{2}p_{i{\bf k}}(t)p_{j{\bf k}}^{\ast}(t).
\end{equation}
In the single--subband case, Eq.\ (\ref{phiij}) takes a
familiar form $\phi_{{\bf k}}(t)=1-|p_{{\bf k}}(t)|^2$ ---
the usual Pauli blocking factor in the coherent limit\cite{pri99}; 
in a multi--subband case, the latter is a matrix.

Eqs.\ (\ref{pol}--\ref{phiij}) are used here to study the pump/probe
signal of the multi-subband QW during 
negative time delays ($\tau<0$) and  for off-resonant excitation
with detuning $\Omega \gtrsim E_F$, in which case 
the coherent effects dominate.
Similar to the single--subband case\cite{pri99},
the above expressions apply for $\mu_i{\cal E}_p/\Omega\lesssim 1$ 
(or $t_{p}$, $\mu_i{\cal E}_pt_p\lesssim 1$ for short pump pulse
duration). For $\Omega \gtrsim E_{M}$ (or for $E_M t_{p} \leq 1$),
$E_{M}\ll E_F$ being the characteristic Coulomb energy of the FS
excitations, the corrections to the above effective parameters 
due to  pair--pair and pair--FS interactions \cite{pri99} can be
neglected  for simplicity since they are perturbative\cite{s-r} in
the screened interactions. One can also show \cite{pri99} 
that, due to the FS Pauli blocking and the screening,
the pump-induced corrections in the interaction
potentials in (\ref{eff1}) are suppressed, as compared to the
self-energies, by a factor $(E_M/\Omega)^2$ 
(or $(E_M t_{p})^{2}$ for short pump duration)
and can therefore be neglected in the excitation regime of interest
here. Finally, similar to the linear absorption
calculations\cite{mue90}, the effects of $V_{ee}$
can be taken into account via a screened {\em e--h} potential  in
$V_{eh}$ and by treating the {\em e-e} scattering within the
dephasing time approximation \cite{pri99}; indeed, for off-resonant
excitation the {\em e-e} scattering is suppressed, while at the same
time, due to high FS electron density, the build--up of
screening in doped QW's occurs during time scales  shorter than the
typical pulse duration  $\sim$ 100\ fs\cite{pri99}.

Thus, in the coherent limit, the effective Hamiltonian (\ref{eff1}) 
has the same operator form as the ``bare'' Hamiltonian (\ref{H0}), but
with time--dependent band dispersions. To calculate the polarization
(\ref{pol}), we adopt the multi-subband generalization of the
coupled cluster expansion method (CCE) for
time--dependent Hamiltonians \cite{arp83,sch78,pri99}.
We consider the physically relevant limit
of large hole mass and include the hole recoil broadening 
in the dephasing time. Under such conditions, the
CCE provides an exact description  of the dynamics arising 
from the effective Hamiltonian (\ref{eff1}) 
and thus accounts for 
the {\em e--h}  correlations leading to the 
unadiabadic response of the FS to the pump pulse
{\em nonperturbatively} [beyond the 
Hartree--Fock approximation (HFA)]\cite{perakis,pri99}.

Our approach has a straightforward physical interpretation. The
photoexcited {\em e--h} state
${\cal K}(t,\tau)\tilde{U}_{i}^{\dag}(\tau)|0 \rangle$,
entering into (\ref{pol}), can be viewed as describing
the propagation of the {\em e--h} pair with amplitude
$\Phi_{ij}({\bf k},t)$ excited by the probe pulse at time $\tau$,
 dressed by the scattering of the FS excitations (dynamical FS response). 
The latter leads to a dynamical broadening
described by the amplitude $s_{ij}({\bf p},{\bf k},t)$ 
that satisfies the differential equation\cite{sch78,pri99} 
\begin{eqnarray}
\label{for s}
i\frac{\partial s_{ij}({\bf p},{\bf k},t)}{\partial t}=
(\epsilon_{i\bf p}^c-\epsilon_{j\bf k}^c)s_{ij}({\bf p},{\bf k},t)
&&
+\sum_{l} [\Delta\epsilon_{il\bf p}^c(t)s_{lj}({\bf p},{\bf k},t)
-\Delta\epsilon_{lj\bf k}^c(t)s_{il}({\bf p},{\bf k},t)]
\nonumber\\&&
-\sum_{l}\tilde{v}_{il}({\bf p},t)[\delta_{lj}
+\sum_{q>k_F}s_{lj}({\bf q},{\bf k},t)],
\end{eqnarray}
with initial condition $ s_{ij}({\bf p},{\bf k},\tau)=0$,
and {\bf p} and {\bf k} labeling respectively the ($i$th
subband) FS electron and the ($j$th subband) FS hole. Since
only the first subband is  occupied, the only non--zero
components of $s_{ij}$ are $s_{11}({\bf p},{\bf k},t)$ and
$s_{21}({\bf p},{\bf k},t)$, which describe the intra and
intersubband FS excitations 
respectively. The photoexcited 
 {\em e--h} pair wavefunction
$\Phi_{ij}({\bf k},t,\tau)$ satisfies the
Wannier--like equation
\begin{eqnarray}
\label{for phi}
i\frac{\partial\Phi_{ij}({\bf k},t)}{\partial t}=
\sum_{l} [\epsilon_{il\bf k}^c(t)
+\delta_{lj}[\epsilon_{\bf k}^v(t)
+\epsilon_{A}(t)]-i\Gamma]\Phi_{lj}({\bf k},t)
-\sum_{l,q>k_F}\tilde{v}_{il}({\bf k},t)\Phi_{lj}({\bf q},t)
\end{eqnarray}
with initial condition
$\Phi_{ij}({\bf k},\tau)=\phi_{ij{\bf k}}(\tau)$, where
$\epsilon_{A}(t)=-\sum_{k'<k_F} [v_{11}
+\sum_{p'>k_F}s_{11}({\bf p}',{\bf k}',t)v_{11}]$
is the self--energy due to the readjustment
of the FS to the photoexcitation of a hole\cite{pri99,sch78} 
and $\Gamma$ is the inverse dephasing time due to 
all the processes not included in $H$.
In Eqs.\ (\ref{for s})\ and\ (\ref{for phi}),
$\tilde{v}_{ij}({\bf k},t) =v_{ij}-
\sum_{l,k'<k_F}s_{il}({\bf k},{\bf k}',t)v_{lj}$
is the effective {\em e--h} potential 
whose time--dependence is due to the 
dynamical FS response \cite{pri99}.
Note that it is the interplay between this effective potential and
the pump-induced self-energies that gives rise to the unadiabatic FS
response to the pump field.
In terms  of $\Phi_{ij}({\bf k},t)$, the polarization
(\ref{pol}) takes the simple form ($t>\tau$)
\begin{equation}
\label{pol1}
P(t)=-i
{\cal E}_{s}e^{-i\omega_pt}
\sum_{ijl}\mu_i\mu_j\sum_{k>k_F}
\Phi_{il}({\bf k},t)\phi_{ij{\bf k}}^{\ast}(t),
\end{equation}
with $\phi_{ij}({\bf k},t)$ given by (\ref{phiij}). The
nonlinear absorption spectrum is then proportional to
${\rm Im}P(\omega)$, where $P(\omega)$ is the Fourier
transform of the rhs of (\ref{pol1}).

{\em Numerical results}.--- Below we present our results for
the evolution of the pump/probe spectra of the FES--exciton
hybrid. The spectra were obtained by the numerical solution 
of the coupled equations  (\ref{for phi})\ and\ (\ref{for s}), with
the time--dependent band dispersions 
$\epsilon_{ij\bf k}^c(t)$ and $\epsilon_{\bf k}^v(t)$.
The calculations were performed at zero temperature for
below--resonant pump  with detuning $\Omega \sim E_F$
and duration $t_pE_F/\hbar=2.0$, and by 
adopting the  typical values of parameters
$v_{12}/v_{11}=0.2$, $\Gamma=0.1E_F$, and 
$v_{11} {\cal N} = 0.3$, ${\cal N}$ being the density of states, 
previously extracted from fits to the linear absorption spectra 
\cite{mue90,mahan}
($E_F\sim 15-20$ meV in typical GaAs/GaAlAs QW's\cite{kim92,bre95}).
Note, however, that similar results were also obtained for a broad 
range of parameter values. In Fig. 1(a) we plot the nonlinear
absorption spectra at different negative
time delays $\tau<0$. For better visibility, the curves are
shifted vertically with decreasing $|\tau|$ (the highest curve
represents the linear absorption spectrum). For the chosen value
of $\Delta$,  the FES and excitonic components of the hybrid are
distinguishable in the linear absorption spectrum, with the FES peak
carrying larger oscillator strength. It can be seen that, at short
$\tau<0$, the oscillator strength is first transferred to the
exciton and then, with further increase in $|\tau|$, back to the FES. At
the same time, both peaks experience a blueshift, which is
larger for the FES than for the exciton peak because 
the ac--Stark effect\cite{ssr86} for the 
exciton is weaker due to the subband separation $\Delta$.

The transient exchange of oscillator strength originates
from the different nature of the 
FES and exciton components of the hybrid. At negative
time delays, the time--evolution of the exciton is
governed by its dephasing time, which is essentially determined
by the homogeneous broadening $\Gamma$ (in doped systems the
exciton--exciton correlations do not play a significant role
due to the screening). The pump pulse first leads to a bleaching of
the exciton peak, which then recovers its strength at
$|\tau|\sim\hbar/\Gamma$. On the other hand, since the FES 
is a many--body {\em continuum} resonance, (i) the bleaching
of the FES peak is stronger, and (ii) the polarization decay
of the FES is determined not by $\Gamma$, but by the scattering
with the low--lying FS excitations. 
This leads to much faster dynamics, roughly determined by the inverse 
Coulomb energy $E_{M}$ \cite{pri99}.
However, the time--evolution of the hybrid spectrum 
is not a simple superposition of the dynamics of its
components. Indeed, the pump-induced self-energies lead 
to the  flattening of the subbands or, to the first
approximation, to a time--dependent increase in the 
effective mass (and hence the density of states), which 
in turn increases the {\em e--h} scattering\cite{pri99}.  
Important is, however, that, due to the subband
separation and different nature of the resonances, such an
increase is stronger for the FES. Therefore, the effect of the pump
is to reduce the excitonic enhancement of the FES peak (coming from
the resonant scattering of the photoexcited electron by the exciton level)
as compared to the linear absorption case, 
resulting in the  oscillator strength transfer from the FES back
to exciton. In fact, such a transfer is strong even for smaller
$\Delta$ [see Fig. 1(b)]. It should be emphasized that the above
feature cannot be captured within the HFA.
Indeed, the latter approximates the FES by a bound 
state\cite{pri99} and thus neglects the difference between the FES
and exciton dynamics originating from the unadiabatic response of
the FS to the change in the {\em e--h} correlations. 
This is demonstrated in Fig. 1(c) where we show the spectra obtained
without the FS dynamical response, i.e., by setting
$s_{ij}=0$. Although in that case both peaks show blue shift and
broadening,  there is no significant transfer of oscillator strength 

In conclusion, we investigated theoretically the coherent nonlinear
optical response of the FES--exciton hybrid in a QW with partially
occupied subbands.  We found a strong redistribution of the
oscillator strength between the FES and exciton peaks
which is caused by the different dynamics of the FES and exciton
components of the hybrid as well as by their coupling due to the 
{\em e--h} correlations. This originates from the dynamical Fermi
sea response and leads to a strong transient changes in the
pump/probe spectra. Such systems can be used to probe the role of
the many--body correlations in the Fermi liquid versus bound states
dynamics.
 
This work was supported by the NSF grant ECS-9703453, 
and by HARL, Hitachi Ltd.  
The work of D.S.C. was supported by the Director, Office
of Energy Research, Office of Basic Energy Sciences, Division
of Material Sciences of the U.S. Department of Energy,
under Contract No.\ DE-AC03-76SF00098.


\begin{figure}
\caption{
(a) Calculated pump/probe spectra with (a) $\Delta=1.7E_F$, 
(b) $\Delta=1.6E_F$, and (c)  $\Delta=1.6E_F$ (HFA), 
for short pump duration $t_pE_F/\hbar=2.0$, and negative time 
delays $\tau\Gamma/\hbar=-2.0$ (lowest curve), $-1.2$,
$-0.6$, $-0.4$, $-0.2$, 0, and linear absorption spectrum (upper curve).
Inset: schematic plot of the energy spectrum of the two-subband
QW (right) and absorption spectrum (left).
}
\end{figure}


\clearpage
\begin{center}
\epsfxsize=4.0in
\epsffile{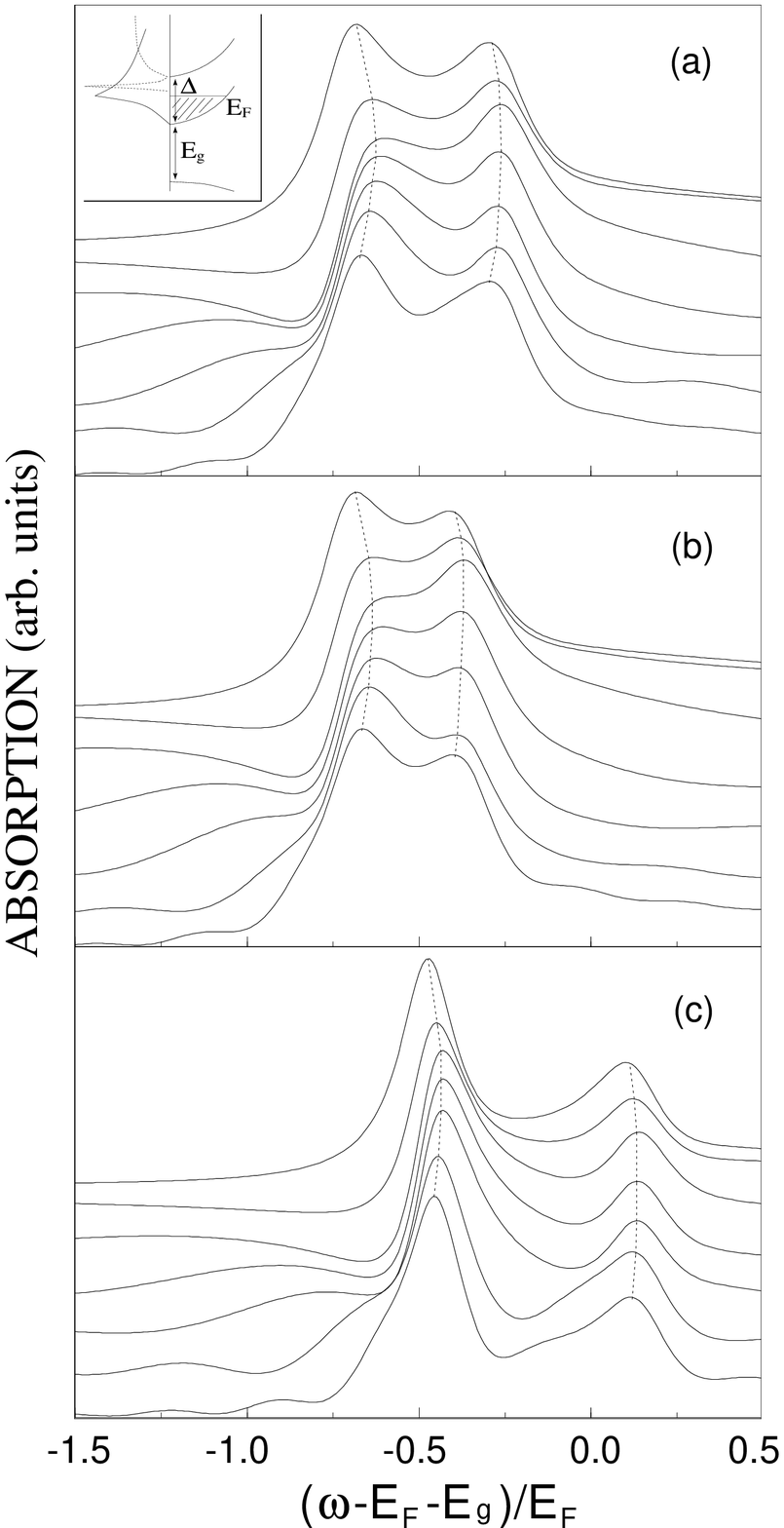}
\end{center}
\vspace{10mm}
\centerline{FIG. 1}
\end{document}